\documentclass[AMA,STIX1COL]{WileyNJD-v2}

\usepackage[colorinlistoftodos, textwidth=13mm, shadow]{todonotes}

\articletype{Article Type}

\received{26 April 2016}
\revised{6 June 2016}
\accepted{6 June 2016}

\begin{document}

\title{Statistical Review of Animal Trials - A Guideline}

\author[1,2]{Sophie K. Piper*}
\author[1,2]{Dario Zocholl}
\author[3]{Ulf Toelch}
\author[1,2,4]{Robert Roehle}
\author[1,2]{Andrea Stroux}
\author[5]{Johanna Hößler}
\author[5]{Anne Zinke}
\author[1,2]{Frank Konietschke}

\authormark{Piper\textsc{et al}}

%Charité – Universitätsmedizin Berlin, corporate member of Freie Universität Berlin and Humboldt-Universität zu Berlin, Institute of Biometry and Clinical Epidemiology, Charitéplatz 1, 10117 Berlin, Germany

%Berlin Institute of Health at Charité – Universitätsmedizin Berlin, Charitéplatz 1, 10117 Berlin, Germany

\address[1]{\orgdiv{Charité – Universitätsmedizin Berlin, corporate member of Freie Universität Berlin and Humboldt-Universität zu Berlin, Institute of Biometry and Clinical Epidemiology },\orgaddress{\state{Charitéplatz 1, 10117 Berlin}, \country{Germany}}}

\address[2]{\orgdiv{Berlin Institute of Health at Charité – Universitätsmedizin Berlin}, \orgaddress{\state{Charitéplatz 1, 10117 Berlin}, \country{Germany}}}

\address[3]{\orgdiv{Berlin Institute of Health at Charité – Universitätsmedizin Berlin, QUEST Center for Responsible Research}, \orgaddress{\state{Anna-Louisa-Karsch Str. 2, 10178 Berlin}, \country{Germany}}}

\address[4]{\orgdiv{Charité – Universitätsmedizin Berlin, corporate member of Freie Universität Berlin and Humboldt-Universität zu Berlin, Clinical Trial Office }, \orgaddress{\state{Charitéplatz 1, 10117 Berlin}, \country{Germany}}}

\address[5]{\orgdiv{Landesamt für Gesundheit und Soziales, Referat für gesundheitlichen Verbraucherschutz},  \orgaddress{\state{Postfach 310929, 10639 Berlin}, \country{Germany}}}

\corres{*Corresponding author, \email{sophie.piper@charite.de}}

\presentaddress{Charitéplatz 1, 10117 Berlin,Germany}

\abstract[Summary]{Any experiment involving living organisms requires justification of the need and moral defensibleness of the study. Statistical planning, design and sample size calculation of the experiment are no less important review criteria than general medical and ethical points to consider. Errors made in the statistical planning and data evaluation phase can have severe consequences on both results and conclusions. They might proliferate and thus impact future trials — an unintended outcome of fundamental research with profound ethical consequences. Unified statistical standards are currently missing for animal review boards in Germany. In order to accompany, we developed a biometric form to be filled and handed in with the proposal at the local authority on animal welfare. It addresses relevant points to consider for biostatistical planning of animal experiments and can help both the applicants and the reviewers in overseeing the entire experiment(s) planned. Furthermore, the form might also aid in meeting the current standards set by the 3+3R’s principle of animal experimentation Replacement, Reduction, Refinement as well as Robustness, Registration and Reporting. The form has already been in use by the local authority of animal welfare in Berlin, Germany. In addition, we provide reference to our user guide giving more detailed explanation and examples for each section of the biometric form.\\
Unifying the set of biostatistical aspects will help both the applicants and the reviewers to equal standards and increase quality of preclinical research projects, also for translational, multicenter, or international studies. }


\maketitle

\section{Introduction}\label{sec: Introduction}
Reviews of any experiment involving living organisms are a necessity. They require justification of  need and moral defensibleness through harm benefit assessment of the study. Beyond medical and ethical considerations, design, sample size calculation and the statistical analysis plan of the experiment constitute important review criteria\cite{kilkenny2009survey}. Up to now, however, trained biostatisticians are still underrepresented members of \S 15 Animal Welfare Act committees. An unpublished national survey revealed that only 8 out of 34 ($\sim 24\%$)\footnote{information for one additional  committee was not available}  active committees for animal research in Germany have  appointed a statistician in 2020. 

In comparison with clinical trials, pre-clinical experiments often split up into several sub-experiments across several model systems to support knowledge claims. Due to the complex study structures statistical evaluation demands expertise. Robust results depend on meaningful and efficient statistical planning  of all sub-experiments.  Here, \textit{efficiency} relates to the trade-off between ethical concerns (reduction of animals) and reliability of the statistical methods (in the sense of uncertainty reduction) \cite{festing2018determining, festing1998reducing}. Often, animal testing is conducted with very small sample sizes, e.g. ten or even less animals per group \cite{bonapersona2021historical_control_data}. On the contrary, many statistical methods require medium or large sample sizes for an accurate type-1 error rate control. In case of (very) small sample sizes and when test assumptions are not met, the test procedure might be liberal and thus might tend to over-reject the null hypothesis.
Errors made in the statistical planning and data evaluation phase can have severe consequences on both results and conclusions \cite{strech20193rs}. They might proliferate and thus impact future trials\textemdash an unintended outcome of fundamental research with profound ethical consequences. Therefore,  animal experiments must be efficient in both medical and statistical ways \cite{GuidelineALS}.  However, the complexity and the exploratory nature of the studies typically makes the statistical planning a rather challenging but nevertheless non-negligible task  \cite{kimmelman_distinguishing_2014, percie2020arrive}.  
Existing and well established clinical trial criteria like randomization, blinding, pre-registration, interim analysis, etc., are also applicable in pre-clinical phases \cite{festing2002guidelines}. 
All in all, statistical aspects are key quality criteria and every Animal Welfare Act committee should be aware of their importance.\\

% Unified biometric criteria might help to equate standards and thus enhance the quality of research projects, in particular when the project will be conducted at more than one center or location (national or international). 
%In order to support statistical planning and ethical review, we developed a biometric form to be filled and handed in with the proposal at the local authority on animal welfare. 
%It lists current state of the art points to consider for biostatistical planning and thus statistical review criteria for animal experiments and might help the applicants and reviewers in overseeing the entire (and often rather complex) experiment(s) planned.

With this article we aim to raise awareness for the need of statistical expertise in Animal Welfare Act committees and introduce a form sheet addressing relevant points to consider for biostatistical planning of animal experiments. The form has already been in use by the local authority of animal welfare in Berlin, Germany, and is  available online (in German).\cite{Formblatt} 
Unifying the set of biostatistical aspects will help both the applicants and the reviewers to equal standards and increase quality of preclinical research projects, also for translational, multicenter, or international studies \cite{festing2002guidelines}.
Furthermore, the form might also aid in meeting the current standards set by the \textbf{3+3R}’s principle of animal experimentation \textbf{R}eplacement, \textbf{R}eduction, \textbf{R}efinement \cite{guhad2005introduction} as well as \textbf{R}obustness, \textbf{R}egistration and \textbf{R}eporting \cite{strech20193rs}.

In the following, all criteria will be explicitly presented and explained in Section~\ref{sec: Criteria}. 
Further, we discuss first experiences from both applicants and reviewers in Section~\ref{sec: Discussion}. In addition, we provide a reference to our user guide giving more detailed explanation and examples for each section of the biometric form.

%\\
%{\color{red}
%%- Notwendigkeit von Biometrie in Tierversuchen \\

%-Ziel einer präklinischen Studie im Vgl zu klinscher Studie\\
%-grundlagenforschung, BEnefit, Harm\\
%-explorative Studie: erhofftes outcome, kriterien für konfirmatorische Folgestudie. minimale Effektstärke\\
%- Link zur Paper Serie Ethikkommisionen (introduction)\\
%- Festing, Michael: review on ... faustformel zu t test (abschnitt 2.4 oder diskussion)\\
%- Reproduzierbarkeit \\
%- Planung von explorativen Studien (Darstellung in unserem Formblatt)\\
%- Notwendigkeit, dass Statistiker in den Ethikkommissionen vertreten sind\\
%-Unsere Erfahrungen\\
%- Charite3R und andere 3R Missionen\\
%-Arrive Guidelines\\

%- pot. citation: Guidelines for preclinical animal research in ALS  /
%MND: A consensus meeting. https://doi.org/10.3109/17482960903545334\\

%-potentially cite "where have all the rodents gone?" \cite{holman2016have} , and "Sequential design paper"? \cite{neumann2017increasing}\\
%-tbd: do we use study, experiment or trial, or all interchangably?  -> stick to study, experiment.
%tbd: biometric/ statistic  biostatistician/statistician?
%tbd: sequential/adaptive design-> misleading to statitiscla design with interrim analysis and tests---->change to hirarchical/ subsquent/logical order of experiments?
%}

\section{Biometric Criteria} \label{sec: Criteria}
As mentioned before, we recall important statistical review criteria in this section. All of them are listed in the current biometric planning form used by the local authority of animal welfare in Berlin, Germany. The layout of the form helps both the applicants and reviewers to summarize the primary goal of the study, its (statistical) planning, sample size calculation and verification in a precise, specific and condense way. In the following, we explain the individual items briefly. 
\begin{enumerate}
\item \textbf{Goal of the (sub)-trial}:	A precise description of the (primary) study goal is key for successful implementation and review. This is reflected in item 13 of the ARRIVE guidelines 2.0: "Objectives: Clearly describe the research question, research objectives and, where appropriate, specific hypotheses being tested." \cite{percie2020arrive}
 At this stage, indicating whether the trial is \textbf{exploratory}  or \textbf{confirmatory} is mandatory.\\
 Whereas exploratory research aims at generating new hypothesis, confirmatory research will test these novel hypotheses in a robust manner. Even though there is a gradient between the two, researchers should be aware that confirmatory studies are characterised by increased reliability and validity of experimental evidence. This is particularly important with regard to the translation of results from animal models into clinical contexts \cite{drude_science_2021}. The distinction between research conducted in exploratory or confirmatory mode has an influence on all biostatistical criteria. One important aspect is the balance between Type I errors (false positives) and Type II errors (false negatives). In many biomedical fields, prior probabilities of a hypothesis being true are low. With low numbers of experimental units (i.e. in most cases animals) in exploratory studies the chances of a Type II error are high. This carries the risk of discarding viable hypotheses too early in a research trajectory. That means stringent null hypothesis testing with a focus on standard p-value thresholds in exploratory research will eventually result in a low positive predictive value \cite{ioannidis_why_2005}. Research conducted in exploratory mode should thus focus on high internal validity by for example reducing risk of bias. Whether an exploratory result is based on a true (i.e. biologically meaningful) effect or is false positive is almost impossible to justify. Evidence in such experiments should be judged on effect sizes and uncertainty rather than stringent p-value thresholds. Confirmatory experiments are then needed to collect robust evidence with regard to a specific hypothesis \cite{kimmelman_distinguishing_2014,mogil_no_2017} .
 
\item \textbf{Primary endpoint} of each sub-experiment:

The primary endpoint is the outcome measure that is used to answer the primary research question and it is also the endpoint the study is powered for. It should be specified precisely with the exact measurement method, the unit of measurement and the specific point in time when the measurement for the primary research question is taken. The latter is especially important if the study has a longitudinal design and observations are made at several time points.

\item \textbf{Description of the study design}: A detailed description of the study design is necessary to characterize  whether it is suited to reach the study goals. Design aspects as well as methods against bias (blinding and randomization) should be taken into account. 
\begin{itemize}
\item \textbf{Design}: A detailed description of the study design is fundamental. A flowchart is a useful and helpful tool to illustrate design and workflow of the trial (e.g. https://www.nc3rs.org.uk/experimental-design-assistant-eda).
\item\textbf{Blinding}: Blinding means that information about treatment allocation is withheld from certain or all investigators involved with the aim of minimizing information bias and enhancing the study quality \cite{hirst2014need_for_randomization}. Applicants should distinguish between blinding for conduct of the experiment, assessment of outcome, and analysis. Procedures should be described in detail as in most cases blinding involves several parties that need to be coordinated. In case blinding is not applicable, further methods against bias must be suggested and justified. 
\item	\textbf{Randomization}: Randomization describes the process of randomly allocating the treatments to the study subjects with the aim of minimizing selection bias and distributing potentially influential parameters evenly across groups.

In pre-clinical research, randomization seems not to be as fully integrated as in clinical phases and its impact is often underestimated. Systematic differences between experimental divisions may induce bias and thus might impact the results and aid false conclusions \cite{hirst2014need_for_randomization}. Furthermore, since animals are typically kept in shared cages, possible cluster effects should be taken into account. This applies also to physiological factors like e.g. weight that may have an influence on the primary outcome. Here, block randomisation is necessary. As for blinding, details of the randomization procedure have to be provided. If randomization is not applicable, further methods against bias and confounding must be provided and justified.   
\end{itemize}	
\item	\textbf{Sample size calculation}: The exact number of experimental units allocated to each group as well as the total number of animals in each experiment should be specified (corresponding to ARRIVE guideline item 2a) \cite{percie2020arrive}. Moreover, details of the sample size calculation should include the following information: 
\begin{itemize}
\item	Confirmatory trial:  A summarizing statement giving 	\textit{ \textbf{(i)}} the name of the statistical test used for sample size calculation, \textit{ \textbf{(ii)}} the chosen significance level $(\alpha)$,\textit{ \textbf{(iii)}} the desired power $(1-\beta)$ with $\beta$ being the type II error-rate, \textit{ \textbf{(iv)}} whether a one- or a two-sided significance-level will be used, and, \textit{ \textbf{(v)}} the  physiologically relevant (minimal) effect size with regard to the primary endpoint that is planned to be confirmed. For\textit{ (v)} it is important to state how this effect size was derived, i.e., which expected mean and standard deviation ($SD$) per group are assumed. If possible, this should be backed up by at least one published reference. Since effect size estimates from exploratory studies are often inflated \cite{colquhoun_investigation_nodate} these should be conservatively evaluated if used for subsequent confirmatory trials. Finally,	\textit{\textbf{(vi)}} the software used to calculate the sample size (including version number) should be given.

\item	Exploratory study/ pilot study or orientation trial/ preliminary technical test: \\
 If possible, a priori sample size calculation should be done giving all  required details \textit{ \textbf{(i-vi)}} stated above for confirmatory trials. If this is not possible, it should at least be explained how the sample size was derived. That is, what confidence interval width for the effect estimate can be achieved with the chosen sample size (e.g. width of the 95\% confidence interval). Justification of the sample size could be based on feasibility within a given time frame and laboratory or on the smallest effect size of interest \cite{festing2018determining}. 
A useful tool to a priori evaluate the achievable evidence for a range of effect sizes is a power curve. 

  In clinical pilot trials that wish to estimate mean and variance of a metric endpoint twelve samples per group have been recommended \cite{julious2005sample}. Furthermore, if the trial aims at estimating an effect size for future confirmatory trials, the width of the two-sided $(1-\alpha)$ Wald-type confidence interval $CI=[Mean \mp \tfrac{1.96}{\sqrt{n}}\cdot SD ]$, which is $2\cdot\tfrac{1.96}{\sqrt{n}}\cdot SD$, is a useful criterion for sample size justification. For example, its width is 1 $SD$ if $n=16$ for any arbitrary endpoint \cite{festing2018determining}. Any value smaller than $n=16$ will result in an interval, which is wider than 1 standard deviation.\\
  \textbf{Caveat}: P-values derived from exploratory experiments are not suited to enable decisions towards further experiments like confirmatory studies. Ideally, go/no-go criteria for engaging in confirmatory studies should be defined a priori.\cite{albers2018power, drude_science_2021}
\end{itemize}

 In addition, irrespective of the exploratory or confirmatory nature of the study \textit{\textbf{(vii)}} the number of required reserve animals or dropouts due to premature death, incorrect interventions, etc. should be given. Here, reporting a dropout rate only is not sufficient but the absolute number of additional animals should be stated explicitly. Of note, the absolute number of dropouts is often falsely calculated. If the estimated dropout rate r, e.g 0.2 corresponding to $20\%$ and the required sample size n resulting from the sample size calculation are given, the correct number of additional animals needed would be $n_{dropout}= round up (\frac{n}{(1-r)} -n) $ or for the total number needed:  $n_{total}=n+n_{dropout}=round up (\frac{n}{(1-r)}) $.

\item \textbf{Statistical Analysis} (e.g.: type of statistical model, consideration of relevant covariates, adjustments for multiple testing, secondary analyses, handling of missing data)

A brief summary of the planned methods of analysis in this (sub-)trial should be given including descriptive statistics. Most importantly, this section should contain the analysis of the primary endpoint, which should be in line with the statistical test used to calculate the sample size. Moreover,  variables with a relevant effect on the endpoint, baseline measurements, confounder, repeated measures or site and cluster effects must be considered. Particularly when small sample sizes preclude inclusion of  aforementioned parameters (overparametrization), the specific model choice should be discussed. It must also be reported how missing data will be dealt with or why no missing data are expected. If several subgroups are being compared, it should be declared if and how the analyses will be adjusted for multiple testing. Further, all secondary analyses should at least be briefly reported. This could be for instance any exploratory subgroup analyses, sensitivity analyses, further statistical modelling and graphical illustrations planned.

\item \textbf{Is there a logical or sequential order of the experiments planned?} (e.g. requirements that have to be fulfilled and consequences on any of the following experiments that can arise) 
\par Often animal studies are planned with a distinct order of single experiments. Results of one experiment can have an impact on subsequent  experiments, e.g. on the dosage applied, the operational setting used, the number of subgroups investigated, or the time point of investigation. It should be stated explicitly if there are any prerequisites from previous (sub-)experiments in this trial that have to be fulfilled in order to start with the present (sub-)experiment. If experiments follow a logical order and/or a sequential or adaptive planning, any conditions that lead to a go/no-go decision must be stated clearly, as well as their impact on further experiments and analyses. Such conditions can, but do not have to be of statistical nature. In some cases, a strict biological/medical justification might be sufficient, for example to stop based on exceeding pre-specified thresholds on established scores, or for instance if in other ways a treatment turns out to be not tolerable. Further, there might be arguments from a design perspective to not perform an experiment: if the experiment is clearly a follow-up on a subsequent experiment, it might make no sense to perform the second experiment if the result of the first was negative. Although this is rarely the case for preclinical research, group sequential testing in the strict statistical sense might be planned \cite{neumann2017increasing}. In this case, a clear reporting of time points and nature of the interim analyses is required, i.e. under which conditions will the experiment be terminated and how do results of this (sub)-experiment affect the rest of the trial or follow-up experiments. To facilitate better understanding by the reviewers a flowchart of the sequence of experiments is recommended.
 
\item	\textbf{Summary of the sample size calculation for each (sub-) trial:}\\
For a detailed overview of the final design, the calculated sample size (including possible dropouts) for each group, assumed effect sizes and/or effects to be estimated are summarized in a table. An example is provided in Table~\ref{Tab: Overview}.

\begin{table}[h!]
\caption{Breakdown of the statistical design, effect sizes and sample size calculations}\label{Tab: Overview}
\begin{tabular}{p{2cm}lp{2cm}lp{2cm}lp{2cm}lp{2cm}lp{2cm}lp{2cm}lp{2cm}l}\hline
Group & Primary  & Expected Effect   & Reference for  & Effect Size & Drop Out  & Sample Size \\
     &  Endpoint      & (e.g., means with ) &  expected effect  & & Rate&(Dropouts\\
          &        & (standard deviations) &    & & &included)\\\hline
\textcolor{gray}{A (control)}& \textcolor{gray}{weight [g]}& \textcolor{gray}{A: mean 100g (SD: 20g)} & \textcolor{gray}{Name et al. (2020)} & \textcolor{gray}{Cohen's d=1 }& \textcolor{gray}{25\%} & \textcolor{gray}{$17/0.75 \approx 23$} \\
\textcolor{gray}{B (treatment)    } & \textcolor{gray}{after 3 weeks} &\textcolor{gray}{B: mean 120g  (SD: 20g)}&       &    &  &\textcolor{gray}{$17/0.75 \approx 23$  }\\\hline
$\vdots$ & $\vdots$ & $\vdots$ & $\vdots$ & $\vdots$ & $\vdots$ & $\vdots$\\\hline 
Total: & & & & & & \textcolor{gray}{46} \\\hline

\end{tabular}
\end{table}

\item \textbf{Signatures}
Statistical expertise during the planning phase of the experiment will enhance the quality of the trial. Having the form signed by a biostatistician involved in the planning likely reduces the number of revisions and inquiries by the local authority of animal welfare. In our form the signature is facultative. 

\item \textbf{Preregistration}
Additionally, though this was not included in the biometric from sheet, we highly recommend to preregister the planned trial. Preregistration is an important step towards reproducible and efficient research. So far it cannot be legally required by the animal welfare authorities at the time of application. In preregistrations, important information about the experimental design and analysis is pre-specified and digitally saved in an appropriate portal. One such portal is operated by the Bundesinstitut für Risikobewertung (Federal Insitute for Risk Assessment) \url{https://www.animalstudyregistry.org/}. Currently, information within a preregistration is under embargo for up to five years and can only be accessed by the investigators and later editors and reviewers. Only after publication the preregistration is accessible for third parties making scooping near impossible. Preregistrations also do not preclude exploratory research but help distinguish confirmatory from exploratory aspects of a study.

\section{Concluding remarks}\label{sec: Discussion}
For clinical trials (including early and later phase), ethical review boards are well established\cite{buchner2019tasks,doppelfeld2019medical}. Thorough reviews include verification of the need of the study, its correct implementation along with proper risk/benefit assessments. Especially the latter involves attesting correct choice and usage of the statistical methods applied for planning and analysis of the data. Whereas ethics commissions for clinical trials often appointed a trained statistician\cite{rauch2020comprehensive} , this is usually not the case for ethics commissions with emphasis on preclinical and animal trials. About $ 80\%$\footnote{with missing information of one federal state} of the current German animal ethic commissions do not involve a biostatistician. In principle, they consist of members appointed by animal welfare organisations and research institutions. Moreover, unified standards are currently missing and guidelines for pre-clinical trials \cite{festing2002guidelines} on statistical planning and reporting have not yet received much attention. On top, review boards work completely independent due to federal regulations without having unified criteria.\cite{jorgensen2021reviewing} In general, missing of the latter complicates the work of reviewers and does not guarantee an unbiased review process upon known criteria for the applicants. The situation particularly impedes planning and conducting of multi-center pre-clinical trials where every involved research institution must apply for approval by their own local authority.

We here introduced a guideline indexing biostatistical criteria that should be reported by applicants to receive ethics approval. It was our initiative in 2018 to collaborate with the local authority of animal welfare in Berlin, Germany, since we identified an urgent need to improve application quality during our consultancies\footnote{In 2020, we consulted 116 applications for animal experiments.} as well as our work for the local ethical review board. We have frequently observed application forms for animal trials that were unclear in design, imprecise in stating the research question and hypotheses, deficient in reporting of statistical planning  and unfortunately also sometimes simply inappropriate by pasting text blocks from former applications. Our biometric form sheet has been implemented by the local authority of animal welfare in March 2019 on a voluntary basis and its use is obligatory since January 2021.  
After two years, we can already summarize that our form sheet helps  reviewers in assessing research goals and trial design also in rather complex pre-clinical trials. Applicants, on the other side, have not always received the new biometric form sheet with enthusiasm.
For those without sound education in statistics it is partly overwhelming in what detail they should now explain their primary research goal, sample size planning and statistical analysis - which was much less strictly handled before. Usually, several hours of consultancy work are needed until the form sheet is ready to be signed by the consulting statistician.  %Zur Info von Uwe: seit 1.1. 2020
%•	116 TVAs beraten (inkl der vorher begonnenen und der noch nicht vollständig abgeschlossenen)
%•	für die seit 1.1. 2020 begonnenen (inkl. der noch nicht abgeschlossenen)
%o	mittlere Gesamtdauer 219,1584 Minuten
%o	mittlere Besprechungsdauer 83,26733 Minuten
%o	mittlere Bearbeitungszeit ohne Kunden 135,8911 Minuten 
 Most often the distinction between exploratory and confirmatory research is not clear to applicants. 
From February to December 2020, roughly 150 application forms for animal experiments have been reviewed by the animal welfare committee of the the local authority of animal welfare in Berlin. Of these applications, about $35\%$ used the new biometric form, which was recommended but not required at that time. If the form was not used, about $ 90\%$ of the applications had statistical deficiencies that demanded major revision. With the new form, deficiencies were less frequent (about $70 \%$), though still far from satisfactory.  We believe this is based on the fact that most applications were completed without the help of a biostatistician. We are optimistic that deficiencies will decrease further in 2021, as the form is now mandatory and early assistance from a biostatistician is strongly recommended. To assist applicants, we have written a constantly updated user guide on how to complete the biometric form. There, we extend the motivation and explanation for each point in the form. Moreover, we provide  explicit examples for an exploratory and a confirmatory study (see doi xxx - to be announced).

In conclusion, preclinical researcher need biostatistical support  when planning their experiments. From our consultancy experiences, we developed and implemented a biometric form sheet to guide applicants. This is a first step to standardize applications and streamline ethical reviews.  Finally, we plan for broader dissemination in collaboration with the remaining animal welfare authorities in Germany.

\end{enumerate}
%\backmatter

\section*{Acknowledgments}
This work was supported by the Deutsche Forschungsgemeinschaft grant number DFG KO 4680/4-1.

\subsection*{Author contributions}
SKP,  DZ, UT, AS, RR, and FK designed the biometric form sheet and prepared parts and critically revised the manuscript. JH and AK acquired background information about the local animal welfare authorities in Germany and carefully revised the manuscript. 

\subsection*{Financial disclosure}

None reported.

\subsection*{Conflict of interest}

The authors declare no potential conflict of interests.

\section*{Data Availability Statement}

Anonymized data about the usage of the biometric form sheet are available at \url{https://doi.org/10.5281/zenodo.5615540}.

%\begin{thebibliography}{99}
%\bibliographystyle{NJDnatbib.sty}
\bibliography{WileyNJD-AMA}
%\bibitem[Bauer and Bauer(1994)Bauer, P. and Bauer, M.M.]{bib1}Bauer, P. and  Bauer, M. M. (1994). Testing equivalence simultaneously for location and  dispersion of two normally distributed populations.  \textit{Biometrical  Journal} \textbf{36}, 643--660.
%\bibitem[Farrington, C. P. and Andrews, N. (2003)]{bib2}Farrington, C.P. and Andrews, N. (2003). Outbreak detection:
%Application to infectious disease surveillance. In: Monitoring the Health of Populations (eds. R. Brookmeyer and D. F. Stroup), Oxford University Press, Oxford,\break 203--231.
%\bibitem[Rencher(1998)Rencher, A.C.]{bib3}Rencher, A. C. (1998).  \textit{Multivariate Statistical Inference and Applications}. Wiley, New  York. 

%\bibitem[Rauch (2020)]{Rauch2020} Rauch, G., Hafermann, L., Mansmann, U.,  Pigeot, I. (2020). Comprehensive survey among statistical members of medical ethics committees in Germany on their personal impression of completeness and correctness of biostatistical aspects of submitted study protocols. \textit{BMJ open} \textbf{10(2)}, 1--10.
%\end{thebibliography}

\appendix

\end{document}